# smFRET reveals DHX36 repetitive binding — not unfolding — of G-quadruplexes

Hai-Lei Guo, Na-Nv Liu, & Xu-Guang Xi

Recently, Chen *et al*.[1] solved the crystal structure of bovine DHX36 bound to a guanine-quadruplex (G4) composed of a *Myc*-promoter-derived G4-forming sequence followed by a 3'-single-standed extension of seven thymidines (G4$^{Myc}$). They observed that the DHX36-specific motif (DSM) binds the top canonical G-tetrad at the 5'-end, which is transformed into a reorganized non-canonical tetrad (G·G·A·T) with a one-base translocation at the 3' end (hereafter, this G4 is called a quasi-G4 (G4$^{Myc\text{-quasi}}$). To confirm that the observed one-base translocation also occurs in solution, they performed single-molecule fluorescence resonance energy transfer (smFRET) assays and demonstrated that the smFRET time trace is oscillatory, characterized by smFRET values of FRET$^{mid}$ (≈ 0.65) and FRET$^{low}$ (≈ 0.40) with an oscillation amplitude of ≈ 0.22~0.24. This finding was taken as strong evidence in support of an ATP-independent repetitive unfolding between the canonical (G4$^{Myc}$) and reorganized G4 (G4$^{Myc\text{-quasi}}$) as observed in the crystal structure. However, the smFRET values recorded with the mutants, which were mainly mutated in DSM, remained at FRET$^{low}$ (≈ 0.40), indicating that mutations do not impair one-base translocations, but block G4 in a one-base translocated state. This conclusion contrasts sharply with their interpretation that DHX36 binding induces one-base translocations: if DHX36 repetitively unfolds G4, impairing DSM binding will affect one-base translocations and smFRET values determined with these mutants should consequently remain at FRET$^{mid}$ (≈ 0.65) (non-translocated state), not at FRET$^{low}$ (≈ 0.40) (translocated state). Furthermore, their smFRET time traces were recorded with only 3'-end Cy3 labeling of G4 DNA. Therefore, other possibilities are conceivable. Establishing a convincing relationship between the recorded oscillatory smFRET time trace and the repetitive unfolding of G4 involves answering at least three fundamental questions. First, does the oscillation phenomenon depend on the Cy3-labeling position? Second, is it possible to experimentally record smFRET values that specifically result from one-base translocations? Third, what are the physical means of the oscillation amplitude, with its low or high values?

If the oscillatory time traces recorded with Cy3-labeled G4$^{Myc}$ at its 3' end reflect repetitive unfolding, then the similar or stronger oscillation signals should be recorded with G4$^{Myc}$ harboring a Cy3 label just upstream of the 3'-G-tetrad (smG4$^{\text{Cy3-T3}}$) (**Fig. 1a**), because this fluorescent labelling position should be even more sensitive to one-base translocations. After having checked that Cy3-labeled smG4$^{Tel\,\text{Cy3-T3}}$ does not affect the binding/unfolding activities (**Extended Data Fig. S1**), our smFRET time traces recorded with these substrates revealed that, although the oscillatory behavior was reproduced with the original 3'-end-Cy3-labeled G4$^{Myc}$ (smG4$^{\text{Cy3-T9}}$) (blue line in **Fig. 1b**), the curve recorded with smG4$^{\text{Cy3-T3}}$ was basically flat (gray line in **Fig. 1b**). These inconsistent results, together with Chen *et al*.'s own observations on DSM mutants, challenge the interpretation that the oscillation "reflects repetitive unfolding between the canonical and reorganized DNA$^{Myc}$ G-quadruplex[1]".

To experimentally and precisely record the smFRET values corresponding to one-base translocations, the best way is to compare the smFRET signal recorded with the canonical G4 (G4$^{Myc}$) and that with the one-base-translocated G4$^{Myc\text{-quasi}}$ structure in the same molecular environment. However, the G4$^{Myc\text{-quasi}}$ observed in the crystal structure cannot be isolated and does not exist as a stable G4 in solution. We therefore designed G4$^{\text{111-sn+1}}$, which is characterized by three G-tetrads linked with one additional base at the 3' end, mimicking the reorganized G4$^{Myc\text{-quasi}}$ (**Fig. 1c, left panel**). Eight pairs of G4 substrates (G4$^{Myc\text{-sn}}$ and G4$^{\text{111-sn+1}}$, where n = 7-15) were prepared, in which G4$^{Myc\text{-sn}}$ corresponds to G4$^{Myc}$ with a 3' single-stranded DNA (ssDNA) of length n and G4$^{\text{111-sn+1}}$ corresponds to G4$^{111}$ with a 3' ssDNA of length n+1 (**Fig. 1c, left panel**). All smFRET time traces recorded with the above-mentioned substrates in the same experimental conditions were oscillatory — as Chen *et al*. had observed — and the FRET$^{\text{mid}}$ and FRET$^{\text{low}}$ values of each trace were explicitly determined (**Fig. 1c, middle panel** and **Extended Data Fig. S2**). Plotting FRET$^{\text{mid}}$ and FRET$^{\text{low}}$ values as a function of the ssDNA length of each G4 substrate revealed that the values (FRET$^{\text{mid}}$ and FRET$^{\text{low}}$) decreased proportionally with the increase in the ssDNA tail length and the amplitudes (FRET$^{\text{mid}}$-FRET$^{\text{low}}$) were invariably equal to 0.20~0.24, whether determined using G4$^{Myc\text{-sn}}$ or G4$^{\text{111-sn+1}}$ (**Fig. 1c, right panel**). The paired comparison of the smFRET$^{\text{mid}}$ value obtained with G4$^{Myc\text{-sn}}$ and G4$^{\text{111-sn+1}}$ at a given n value gave a statistically significant constant differential value ($\Delta$FRET$^{\text{mid}}$) of ≈ 0.02~0.03 (**Fig. 1c, right panel**). $\Delta$FRET$^{\text{low}}$ showed the same pattern. Finally, averaging the difference of smFRET$^{\text{mid}}$ (smFRET$^{\text{mid}}$ determined with 7 nt subtracted from smFRET$^{\text{mid}}$ determined with 15 nt) over the 8 ssDNA bases (the 3'-tailed ssDNA of G4$^{\text{Myc}}$ and G4$^{111}$ increased from 7 to 15 and from 8 to 16, respectively) also yielded a value of 0.02~0.03. Analysis of smFRET$^{\text{low}}$ showed strikingly similar results. Thus, both vertical (paired comparison) and horizontal (averaging smFRET values) analysis indicated that it is the value of 0.02~0.03, not the amplitude of 0.22, that correspond to the one-base translocation.

Regarding the value of 0.22, the most reasonable interpretation is that the amplitude of 0.22 does not reflect the unfolding/folding G4 with one-base translocations, but reflects the fluctuation of DHX36 between the 3'- and 5'-G-tetrad, with DMS binding to the 3'- or 5'-G-tetrad, giving an oscillatory curve. If this is true, the oscillation amplitude should be proportional to the number of G-tetrads. Four parallel G4s harboring G-tetrad numbers of 2, 3, 4 and 6 were constructed according to the previous studies[2-4] and used for smFRET assays under the same experimental conditions in the absence of ATP. smFRET time traces were oscillatory and the amplitudes were explicitly determined (**Fig. 2a**). **Fig. 2b** demonstrates that there is a strong correlation between the determined amplitude and number of G-tetrads, further supporting the interpretation that the oscillation results from DHX36 swinging back and forth between the 3'- and 5'-G-tetrads of the G4, but is unrelated to one-base translocations. We further reasoned that if the above G4s bearing different G-tetrads undergo one base translocation upon DHX36 binding, and then difference values in smFRET between the canonical and the transformed G4s with one base addition towards 3'-end should be within range of 0.03, rather than in the range of amplitude variation. In accordance with our interpretation, while the tetrads number increased from 2-4, $\Delta smFRET^{mid}$ increase from 0.030-0.035 (**Fig. 2b**).

To determine which smFRET values ($FRET^{mid}$ or $FRET^{low}$) correspond to the binding of DSM on the 3'- or 5'-G-tetrad, we performed smFRET assays with two types of G4. In nonparallel human telomeric G4 (G4-I, prepared with annealing buffer containing sodium), the 5'-G-tretrad is covered by a TTA loop, which likely sterically hinders DSM binding, the smFRET value was exclusively that of DHX36 binding to the 3'-G-tetrad (**Fig. 2c**). In a modified G4 bearing a guanine vacancy in the 3'-G-tetrad (G4-II), which reduces or impairs DSM binding to the 3'-G-tetrad, the smFRET value corresponded to that of DSM binding specifically to the 5'-G-tetrad (**Fig. 2c**). As expected, smFRET values recorded with G4-I and G4-II remained at the $FRET^{low}$ ($\approx$ 0.40) or $FRET^{mid}$ ($\approx$ 0.65) positions, respectively (**Fig. 2c**).

In conclusion, our data challenge Chen *et al*.'s interpretation of smFRET results, i.e. the repetitive unfolding of G4 with one-base translocations. We believe that the observed oscillatory curve represents the alternate binding of DHX36 to the 3'- and 5'-G-tetrad of G4s, rather than the repetitive unfolding between the canonical and the transformed non-canonical G4s. Noteworthily, our results reported here also call into question the previously published smFRET data of helicase-mediated G4 unfolding[5]. Therefore, discriminating the smFRET signal of repetitive binding from that of unfolding/unwinding is very important to avoid misinterpreting smFRET results.

## Methods

**DNA Substrate Preparation**. All DNA substrates used in the study are listed in Extended Data Table S1. Unless otherwise specified, the synthetic G4 DNA was dissolved in annealing buffer containing 20 mM Tris/HCl buffer (pH 7.5) with 100 mM KCl heated to 95 ℃, and allowed to cool slowly to room temperature in a water bath. For the sm6G4 DNA construct, enzyme digestion (*Sac*II or *Nar*I) was carried out after annealing, and the two enzyme-digested products were incubate together for 30 min at 35 ℃.

**Buffers**. The binding/unwinding buffer contained 20 mM Tris/HCl, pH 7.5, 50 mM KCl, 2 mM $MgCl_2$. For single-molecule measurements, 0.8% D-glucose, 1 mg/mL glucose oxidase (266-600 units/g; Sigma-Aldrich, St Louis, MO, USA), 0.4 mg/mL catalase (2000–5000 units/mg; Sigma-Aldrich) and 1 mM Trolox (Sigma-Aldrich) were added to the reaction buffer.

**Single-molecule FRET and stopped-flow assays**. The smFRET and fluorescence-based stopped-flow assays were performed as described in reference[6]. 30 nM DHX36 helicase in the absence of ATP was used in smFRET assay, and stopped-flow assay was performed with 20 nM DHX36 helicase, 4 nM DNA substrate and 1 mM ATP.

## Author affiliations

*College of Life Sciences, Northwest A&F University, Yangling, 712100, China*

Hai-Lei Guo, Na-Nv Liu

*Laboratoire de Biologie et de Pharmacologie Appliquee (LBPA), UMR 8113 CNRS,* ´

*Institut D'Alembert, Ecole Normale Supérieure Paris-Saclay, Université Paris-Saclay, 4, Avenue des Sciences, 91190 Gif-sur-Yvette, France*

Xu-Guang Xi

**Data availability** Data are available from the corresponding author upon request.

**Author Contributions** H.L.G., N.N.L. performed the experiments and/or analyzed the data; and X.G.X. wrote the manuscript and conceived the study.

**Competing interests** Declared none.

**Corresponding author** Correspondence to Xu-Guang Xi.

## Figures

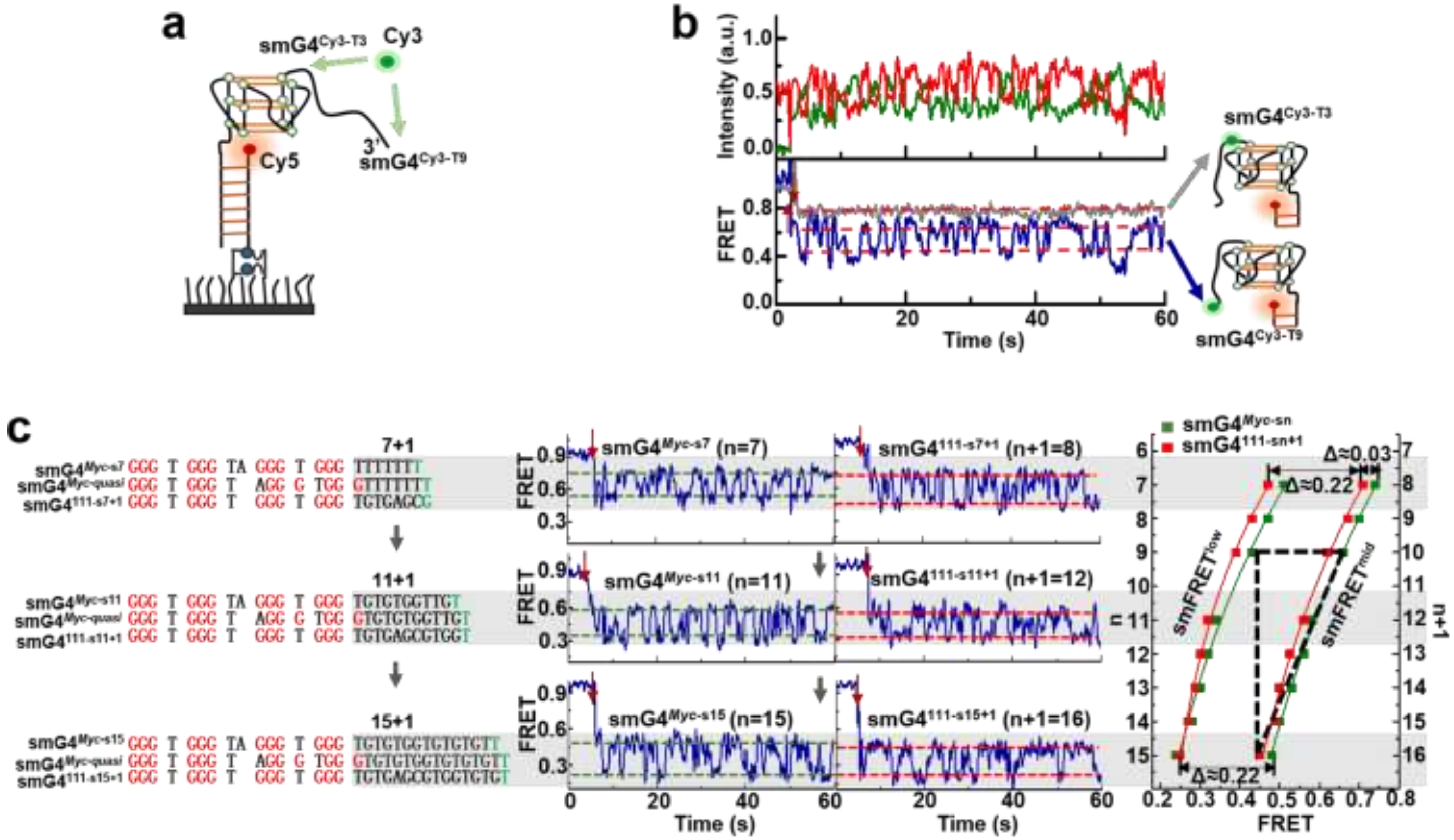

**Figure 1: The oscillatory smFRET curve is best explained by repetitive binding, not by repetitive unfolding with one-base translocations.** **a,** Schematic presentation of labeling G4$^{Myc}$ with a Cy3 label just upstream of the 3'-G-tetrad (smG4$^{\text{Cy3-T3}}$) or at the 3' end of the 9nt ssDNA (smG4$^{\text{Cy3-T9}}$). **b,** Oscillation results from the back-and-forth swinging of the 3'-end-Cy3-labeled 3'-G-tetrad. **c,** DNA sequences used for constructing the canonical G4 (G4$^{Myc\text{-sn}}$, n = 7-15) and the stable reorganized G4 (G4$^{\text{111-sn+1}}$, n = 7-15) with a one-base translocation (left panel); several typical smFRET traces recorded with canonical and reorganized G4s with one additional base at the 3' end (middle panel); smFRET variation as function of the length of G4 harboring ssDNA (right panel).

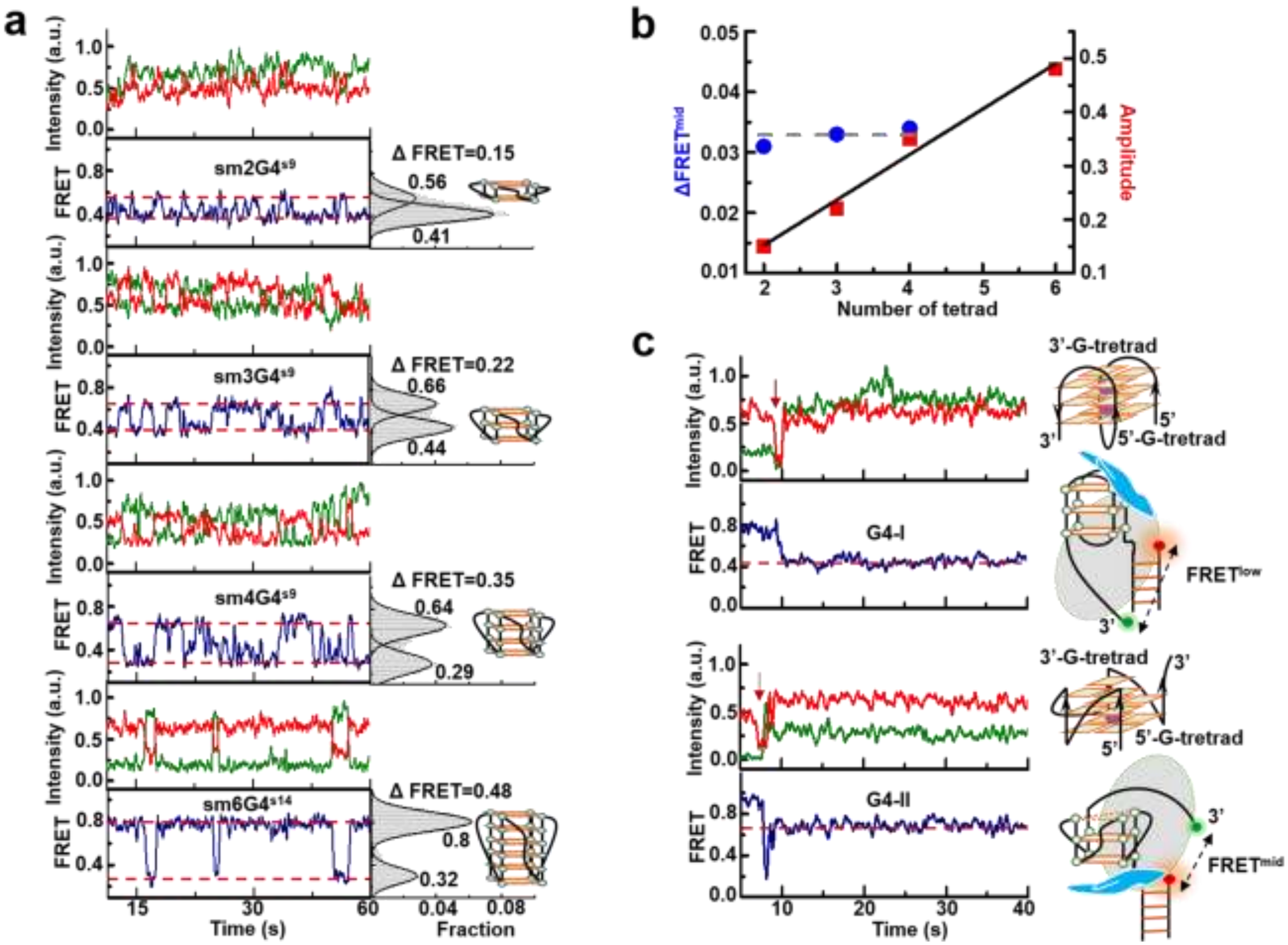

**Figure 2: The correlation between oscillation amplitude and the number of tetrads. a,** Time traces and determined amplitude values with different numbers of tetrads. **b,** The strong correlation between amplitude and the number of tetrads (red square) and the invariant differenced smFRET values between the canonical and the reorganized G4 with different tetrads (blue circle). **c,** The middle and low smFRET values (≈ 0.65 and ≈ 0.40, respectively) correspond to DSM (represented by a hand symbol) binding alternately to the 5'- and 3'-G-tetrads. The oligonucleotides used for the above assays are given in Extended Data Table S1.

## Supplementary Figures and Table

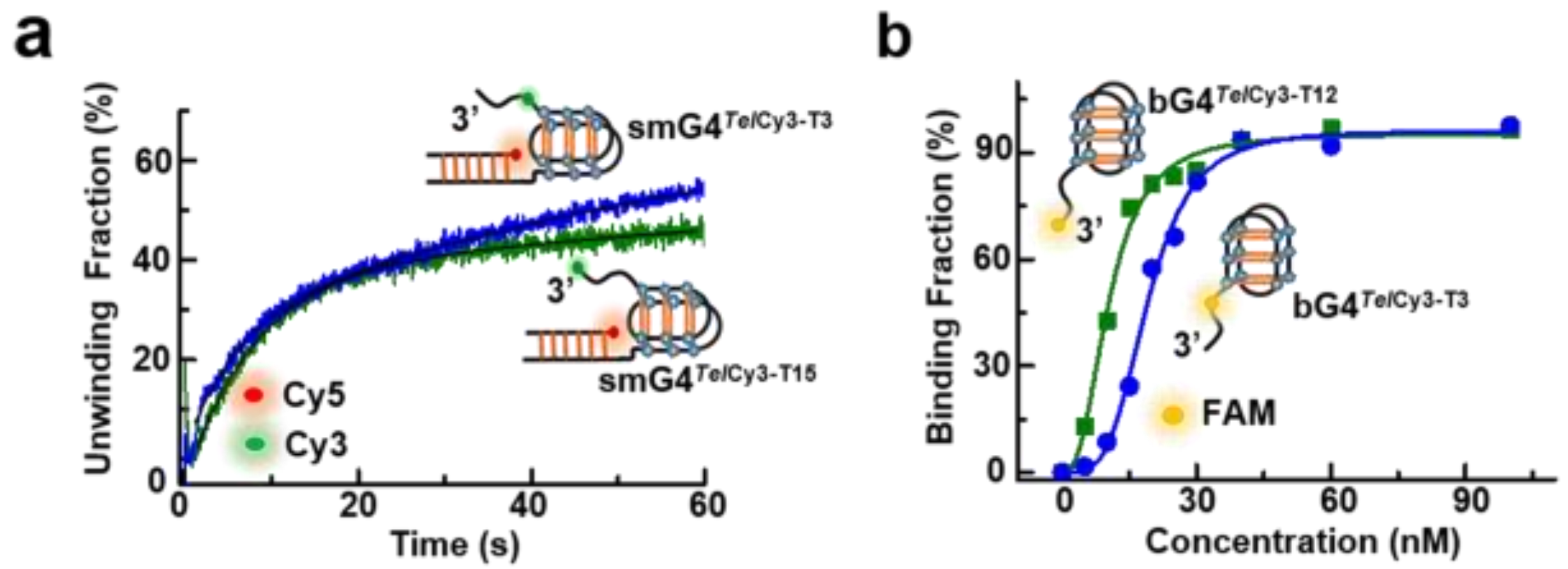

**Extended Data Figure S1:** Cy3 labeling upstream of the 3'-G-tetrad does not impair DHX36 unwinding or binding abilities. The substrates used are schematically presented in **a** and **b**.

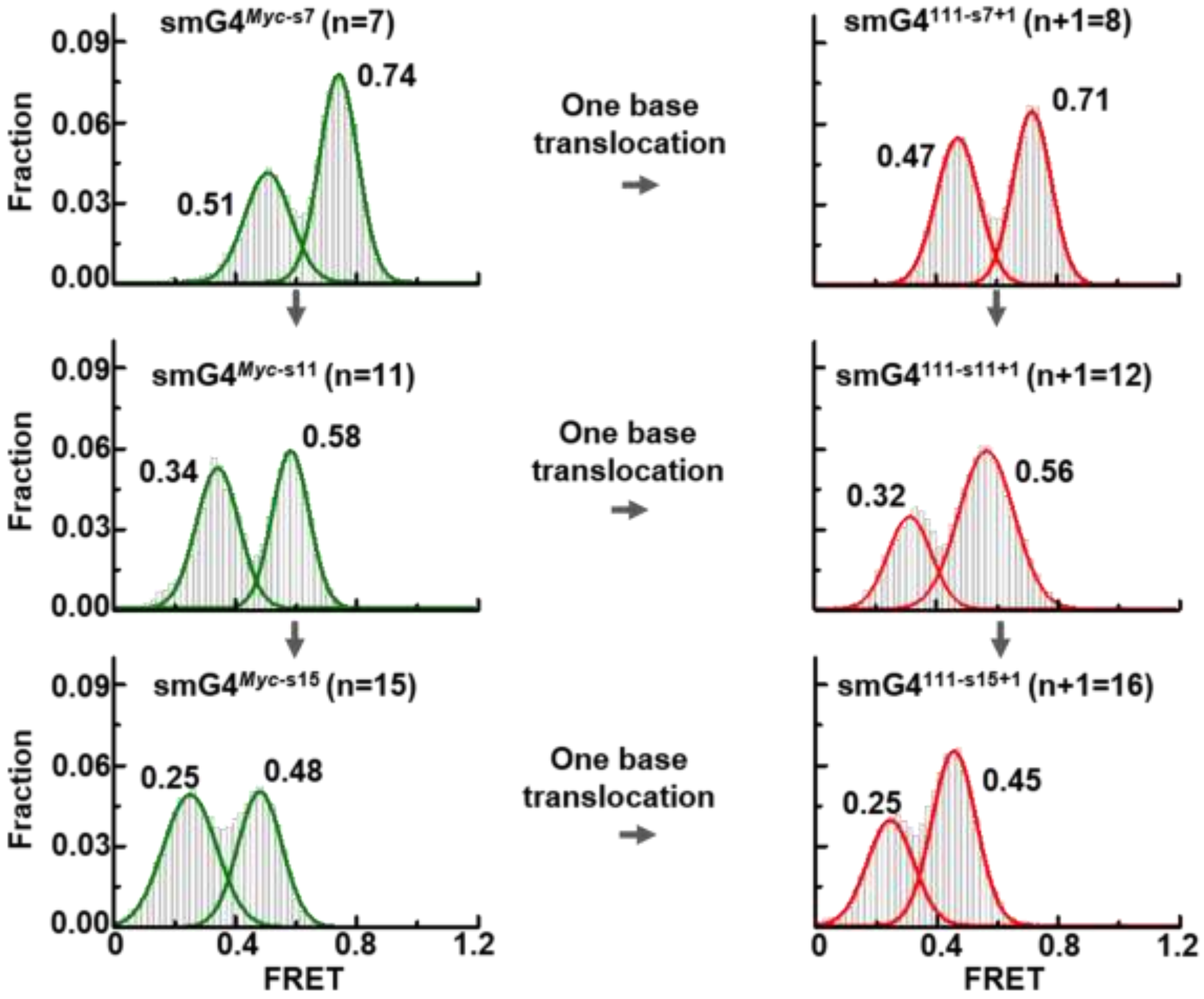

**Extended Data Figure S2:** Typical smFRET histograms of G4$^{Myc\text{-sn}}$ and G4$^{111\text{-sn+1}}$.

## Extended Data Table S1

| Assay | Substrate | Sequence (5'-3')[a] |
|---|---|---|
| smFRET | smG4$^{Cy3\text{-}T3}$ | *TGGCACGTCGAGCAGAGTT*GGGTGGGTAGGGTGGGTG/iCy3dT/GGTGTT |
| | smG4$^{Cy3\text{-}T9}$ | *TGGCACGTCGAGCAGAGTT*GGGTGGGTAGGGTGGGTGTGGTGTT(Cy3) |
| | smG4$^{Tel Cy3\text{-}T3}$ | *TGGCACGTCGAGCAGAGTT*GGGTTAGGGTTAGGGTTAGGGTG/iCy3dT/TGTTGTTGTTGT |
| | smG4$^{Tel Cy3\text{-}T15}$ | *TGGCACGTCGAGCAGAGTT*GGGTTAGGGTTAGGGTTAGGGTGTTGTTGTTGTTGT(Cy3) |
| | smG4$^{Myc\text{-}s7}$ | *TGGCACGTCGAGCAGAGTT*GGGTGGGTAGGGTGGGTTTTTTT(Cy3) |
| | smG4$^{111\text{-}s7+1}$ | *TGGCACGTCGAGCAGAGTT*GGGTGGGTGGGTGGGTGTGAGCG(Cy3) |
| | smG4$^{Myc\text{-}s8}$ | *TGGCACGTCGAGCAGAGTT*GGGTGGGTAGGGTGGGTGTGTGGT(Cy3) |
| | smG4$^{111\text{-}s8+1}$ | *TGGCACGTCGAGCAGAGTT*GGGTGGGTGGGTGGGTGTGAGTGT(Cy3) |
| | smG4$^{Myc\text{-}s9}$(sm3G4$^{s9}$) | *TGGCACGTCGAGCAGAGTT*GGGTGGGTAGGGTGGGTGTGTGGTT(Cy3) |
| | smG4$^{111\text{-}s9+1}$ | *TGGCACGTCGAGCAGAGTT*GGGTGGGTGGGTGGGTGTGAGTGTT(Cy3) |
| | smG4$^{Myc\text{-}s11}$ | *TGGCACGTCGAGCAGAGTT*GGGTGGGTAGGGTGGGTGTGTGGTTGT(Cy3) |
| | smG4$^{111\text{-}s11+1}$ | *TGGCACGTCGAGCAGAGTT*GGGTGGGTGGGTGGGTGTGAGCGTGGT(Cy3) |
| | smG4$^{Myc\text{-}s12}$ | *TGGCACGTCGAGCAGAGTT*GGGTGGGTAGGGTGGGTGTGTGGTTGTG(Cy3) |
| | smG4$^{111\text{-}s12+1}$ | *TGGCACGTCGAGCAGAGTT*GGGTGGGTGGGTGGGTGTGAGCGTGGTG(Cy3) |
| | smG4$^{Myc\text{-}s13}$ | *TGGCACGTCGAGCAGAGTT*GGGTGGGTAGGGTGGGTGTGTGGTTGTGT(Cy3) |
| | smG4$^{111\text{-}s13+1}$ | *TGGCACGTCGAGCAGAGTT*GGGTGGGTGGGTGGGTGTGAGCGTGGTGT(Cy3) |
| | smG4$^{Myc\text{-}s14}$ | *TGGCACGTCGAGCAGAGTT*GGGTGGGTAGGGTGGGTGTGTGGTTGTGTG(Cy3) |
| | smG4$^{111\text{-}s14+1}$ | *TGGCACGTCGAGCAGAGTT*GGGTGGGTGGGTGGGTGTGAGCGTGGTGTG(Cy3) |
| | smG4$^{Myc\text{-}s15}$ | *TGGCACGTCGAGCAGAGTT*GGGTGGGTAGGGTGGGTGTGTGGTGTGTGTT(Cy3) |
| | smG4$^{111\text{-}s15+1}$ | *TGGCACGTCGAGCAGAGTT*GGGTGGGTGGGTGGGTGTGAGCGTGGTGTGT(Cy3) |
| | sm2G4$^{s8}$ | *TGGCACGTCGAGCAGAGTT*GGTGGTGGTGGTGTGAGTG(Cy3) |
| | sm2G4$^{s9}$ | *TGGCACGTCGAGCAGAGTT*GGTGGTGGTGGTGTGAGTGT(Cy3) |
| | sm4G4$^{s8}$ | *TGGCACGTCGAGCAGAGTT*GGTGGTGGTGGTTGTGGTGGTGGTTGTGTGAGTG(Cy3) |
| | sm4G4$^{s9}$ | *TGGCACGTCGAGCAGAGTT*GGTGGTGGTGGTTGTGGTGGTGGTTGTGTGAGTGT(Cy3) |
| | sm6G4$^{s14}$ | GAATTCGGATCCCCGCGG GGGTGGGTGGGTGTTGTTGTTGTGTGT(Cy3) |
| | | CCGCGGGGATCCGAATTC |
| | | *TGGCACGTCGAGCAGATTGTGGGTGGGTGGG GGCGCCGAATTCGGATCC* |
| | | GGATCCGAATTCGGCGCC |
| | G4-I | *TGGCACGTCGAGCAGAGTT*GGGTTAGGGTTAGGGTTAGGGTGTGTGGTT(Cy3) |
| | G4-II | *TGGCACGTCGAGCAGAGTT*GGGTGGGTAGGGTGGTTGTGTGGTT(Cy3) |
| | stem | (Cy5)*ACTCTGCTCGACGTGCCA*-biotin |
| Binding | bG4$^{Tel\ Cy3\text{-}T3}$ | GGGTTAGGGTTAGGGTTAGGGTT/iFdT/TTTTTTTTT[b] |
| | bG4$^{Tel\ Cy3\text{-}T12}$ | GGGTTAGGGTTAGGGTTAGGGTTTTTTTTTTTT(F) |

[a] The G4-forming guanines are shown in red. The complementary sequences are in italics.

[b] F, FAM.